\documentclass[pre,twocolumn]{revtex4-2}
\usepackage{amsmath}
\usepackage{amssymb}
\usepackage{graphicx}
\usepackage{braket}
\usepackage{stmaryrd}
\usepackage{hyperref}
\usepackage{xcolor}

\definecolor{myblue}{RGB}{0,0,255}
\hypersetup{
    colorlinks,
    citecolor=myblue,
    linkcolor=myblue,
    urlcolor=myblue
}

\newcommand{\sqrbr}[1]{\llbracket #1\rrbracket}

\newcommand{\inputUoutput}{(5)}

\begin{document}
\title{Thermodynamic uncertainty relation for quantum first passage process}
\author{Yoshihiko Hasegawa}
\email{hasegawa@biom.t.u-tokyo.ac.jp}
\affiliation{Department of Information and Communication Engineering, Graduate
School of Information Science and Technology, The University of Tokyo,
Tokyo 113-8656, Japan}
\date{\today}
\begin{abstract}
We derive a thermodynamic uncertainty relation for first passage processes in quantum Markov chains. We consider first passage processes that stop after a fixed number of jumps, which contrasts with typical quantum Markov chains which end at a fixed time. 
We obtain bounds for the observables of the first passage processes in quantum Markov chains by the Loschmidt echo, which quantifies the extent of irreversibility in quantum many-body systems. Considering a particular case, we show that the lower bound corresponds to the quantum Fisher information, which plays a fundamental role in uncertainty relations in quantum systems. Moreover, considering classical dynamics, our bound reduces to a thermodynamic uncertainty relation for classical first passage processes. 
\end{abstract}
\maketitle

\section{Introduction}

The thermodynamic uncertainty relation (TUR) provides a fundamental limit of thermodynamic 
machines. It states that the precision of a machine, which is quantified by fluctuations of
thermodynamic currents, is bounded from below by the thermodynamic cost, such as entropy production and dynamical activity.
TURs have been derived for classical systems \cite{Barato:2015:UncRel,Gingrich:2016:TUP,Pietzonka:2016:Bound,Horowitz:2017:TUR,Pigolotti:2017:EP,Garrahan:2017:TUR,Dechant:2018:TUR,Barato:2018:PeriodicTUR,Terlizzi:2019:KUR,Hasegawa:2019:CRI,Hasegawa:2019:FTUR,Vu:2019:UTURPRE,Vu:2020:TURProtocolPRE,Dechant:2020:FRIPNAS,Vo:2020:TURCSLPRE,Koyuk:2020:TUR,Dechant:2020:ContReversal} and quantum systems \cite{Erker:2017:QClockTUR,Brandner:2018:Transport,Carollo:2019:QuantumLDP,Liu:2019:QTUR,Guarnieri:2019:QTURPRR,Saryal:2019:TUR,Hasegawa:2020:QTURPRL,Friedman:2020:AtomicTURPRB,Hasegawa:2020:TUROQS,Sacchi:2021:BosonicTUR,Kalaee:2021:QTUR}. 
Recently, TURs have become a central topic in nonequilibrium thermodynamics.
Besides their theoretical significance, TURs have practical advantages and are used in estimating the entropy production of thermodynamic machines solely from their stochastic trajectories \cite{Li:2019:EPInference,Manikandan:2019:InferEPPRL,Vu:2020:EPInferPRE,Otsubo:2020:EPInferPRE}.

The first passage process is a stochastic process and has been extensively studied in various fields ranging from physics to finance \cite{Redner:2001:FPTbook,Ding:2004:FPTChapter}.
Recently, it is attracting much attention in quantum dynamics as well \cite{Friedman:2017:QFPT,Yin:2019:QFPT}.  
Conventionally, we consider the dynamics of stochastic processes that starts at $0$ and ends at a fixed time $\tau$ ($\tau > 0$). However, in first passage processes, the dynamics stops when the system satisfies some predefined conditions, and therefore the end time of the dynamics, known as the first passage time, is considered a random variable. The predefined conditions could be the system reaching some practically meaningful states, e.g., absorbing states such as extinction, or the number of jumps surpassing a threshold. 
Recently, first passage processes have become increasingly important in stochastic thermodynamics \cite{Saito:2016:EntFluc,Neri:2017:StopTimeST,Neri:2020:StopTime}. For instance, it is possible to extract work by a gambling demon, which monitors its state and determines when to stop the dynamics \cite{Manzano:2021:GambDemon}. 
For TURs in first passage processes, 
relations for classical stochastic thermodynamics were derived in Refs.~\cite{Garrahan:2017:TUR,Pal:2021:FPTTURPRR,Hiura:2021:KUR}. These relations show that the fluctuations of the first passage time are bounded from below by dynamical activity or entropy production. 
TURs for first passage processes are particularly important in biochemical clocks \cite{Barato:2015:UncRel,Barato:2016:BrowClo}. The precision of such clocks is
ideally evaluated through the first passage time for them to accomplish a single chemical reaction cycle.
Although various studies on the first passage problem have been conducted in classical stochastic thermodynamics, 
studies of quantum variants are in a very early stage. 
It was recently proposed that quantum clocks can be implemented by quantum heat engines \cite{Erker:2017:QClockTUR,Mitchison:2019:QMachine}, 
which strongly demands a TUR for the quantum first passage processes.

In this manuscript, we consider a TUR for quantum Markov chains that stop after a fixed number of jumps, which is a particular case of first passage processes. Using the techniques developed in Ref.~\cite{Hasegawa:2021:QTURLEPRL}, we obtain a TUR for the first passage time in quantum Markov chains, whose lower bound comprises the Loschmidt echo. 
The obtained bound concerns two dynamics: the original and perturbed dynamics. 
When the perturbed dynamics is identical to the original dynamics, except that the time scale of the perturbed dynamics is slightly faster or slower than that of the original dynamics, the Loschmidt echo reduces to the quantum Fisher information. This information plays a fundamental role in
several uncertainty relations in quantum systems. 
Considering the classical limit of the derived TUR, we show that our relation reduces to a classical TUR for the classical first passage processes derived so far. 
Considering a two-level atom driven by a classical laser field as an application, 
we show that the fluctuations of the first passage time for a quantum Markov chain become smaller than its classical counterpart.

\begin{figure}
\includegraphics[width=8.5cm]{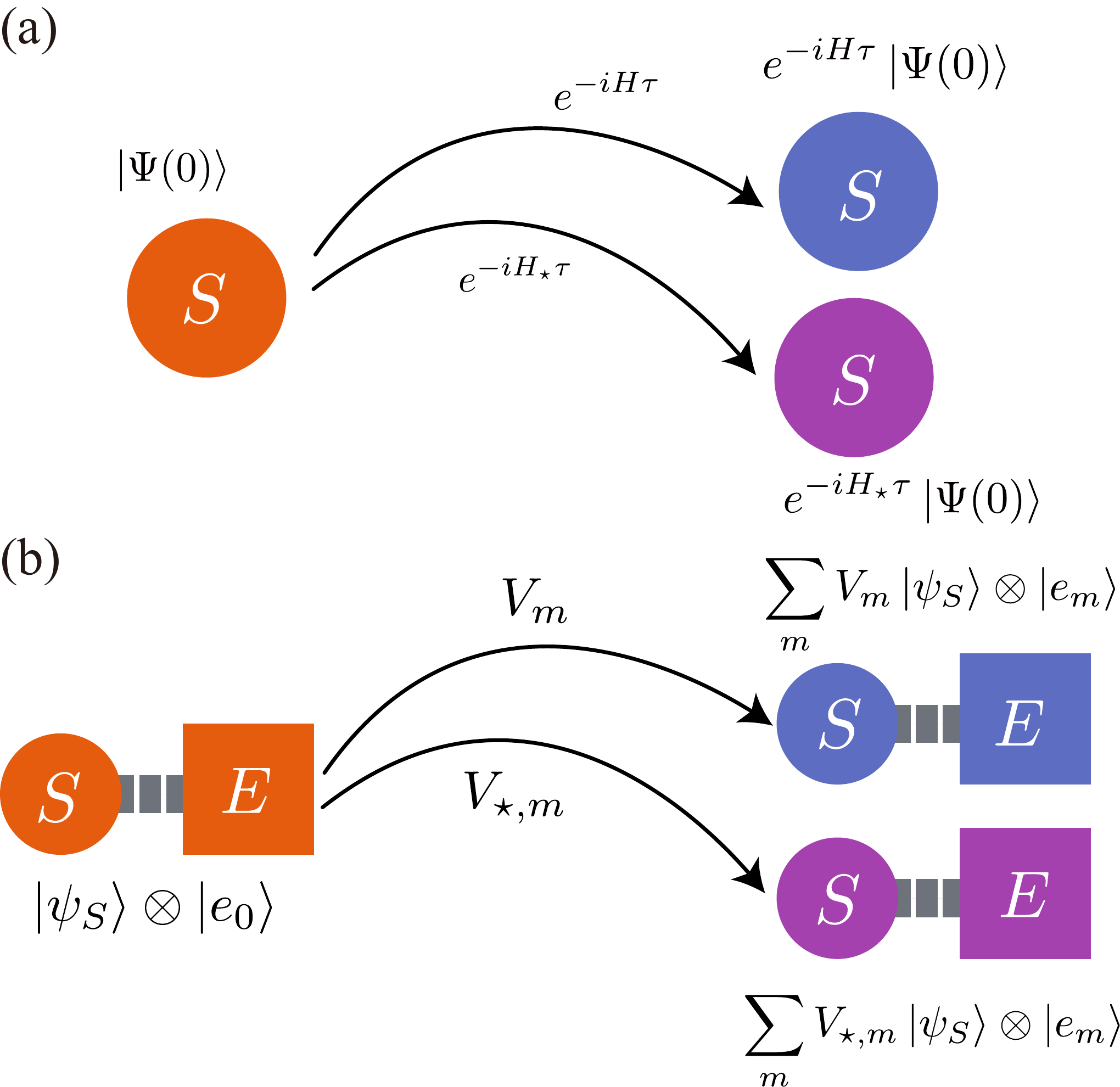} \caption{
Loschmidt echo for (a) closed and (b) open dynamics. $S$ represents the system and $E$ represents the environment. (a)
Loschmidt echo defined by the fidelity between $e^{-iH\tau} \ket{\Psi(0)}$ and $e^{-iH_\star\tau} \ket{\Psi(0)}$ at $t=\tau$, where $H$ and $H_\star$ are the original and perturbed Hamiltonians, respectively. 
(b) Loschmidt echo defined by the fidelity between $\sum_m V_m \ket{\psi_S}\otimes \ket{e_m}$ and 
$\sum_m V_{\star,m} \ket{\psi_S}\otimes \ket{e_m}$, where $V_m$ and $V_{\star,m}$ are the Kraus operators of the original and perturbed dynamics, respectively. 
\label{fig:LoschmidtEcho_PRE}}
\end{figure}

\section{Methods}

The Loschmidt echo is an indicator of quantum chaos in many-body systems.
Let $H$ be a Hamiltonian and let $H_\star$ be a perturbed Hamiltonian. 
When the system is closed, given the initial state $\ket{\Psi(0)}$, the Loschmidt echo $\eta$ is defined by
\begin{equation}
\eta \equiv |\braket{\Psi(0)|e^{iH_\star \tau} e^{-iH\tau}|\Psi(0)}|^2.
\label{eq:Loschmidt_PRE_def}
\end{equation}The Loschmidt echo $\eta$ defined in Eq.~\eqref{eq:Loschmidt_PRE_def} can be regarded as the fidelity between $ e^{-iH\tau}\ket{\Psi(0)}$ and $e^{-iH_\star \tau}\ket{\Psi(0)}$,
which are time-evolved states at $\tau$ induced by $H$ and $H_\star$, respectively, as shown in Fig.~\ref{fig:LoschmidtEcho_PRE}(a). 
If the system is sensitive to the perturbation, the fidelity decays quickly to $0$,
indicating that the Loschmidt echo is conceptually similar to the Lyapunov exponent. 
Because the highly susceptible fidelity under a small perturbation is related to the irreversibility of the dynamics,
the Loschmidt echo can be used to quantify the irreversibility. 

The Loschmidt echo in Eq.~\eqref{eq:Loschmidt_PRE_def} is for closed quantum systems. 
However, we can define the Loschmidt echo for open quantum systems. 
Let us consider a Kraus representation:
\begin{equation}
\rho_S \to \sum_m V_m \rho_S V_m^\dagger,
\label{eq:Kraus_repr}
\end{equation}where $\rho_S$ is a density operator and $V_m$ is a Kraus operator satisfying a completeness relation $\sum_m V_m^\dagger V_m = \mathbb{I}_S$
with $\mathbb{I}_S$ being the identity operator in $S$. This time evolution induced by the Kraus representation can be expressed by a unitary time evolution in a larger space comprising the system $S$ and environment $E$:
\begin{equation}
U_{SE} \ket{\psi_S}\otimes\ket{e_0}=\sum_m V_m \ket{\psi_S}\otimes\ket{e_m},
\label{eq:unitary_Kraus}
\end{equation}where $\ket{e_m}$ constitutes an orthonormal basis of $E$, $\ket{\psi_S}$ and $\ket{e_0}$ are initial states of $S$ and $E$, respectively, and $U_{SE}$ is a unitary operator acting on $S+E$. 
Tracing out $E$ in Eq.~\eqref{eq:unitary_Kraus}, we recover the original Kraus representation of Eq.~\eqref{eq:Kraus_repr}. 
Suppose that a Kraus operator for the perturbed dynamics is $V_{\star,m}$. 
Then the fidelity can be calculated between $\sum_m V_m \ket{\psi_S}\otimes\ket{e_m}$ and
$\sum_m V_{\star,m} \ket{\psi_S}\otimes\ket{e_m}$, as shown in Fig.~\ref{fig:LoschmidtEcho_PRE}(b). Throughout this manuscript, we use $\star$ in subscripts to express the perturbed dynamics. 

We can derive an uncertainty relation from the Loschmidt echo. 
Using the lower bound for the Hellinger distance \cite{Nishiyama:2020:HellingerBound},
we obtain the bounds for observables and the Loschmidt echo  \cite{Hasegawa:2021:QTURLEPRL}. 
Let $\ket{\Psi}$ and $\ket{\Psi_\star}$ be two pure states and let $\mathcal{F}$ be an Hermitian observable that is applied to $\ket{\Psi}$ or $\ket{\Psi_\star}$. 
We define the mean and standard deviation of $\mathcal{F}$ by $\braket{\mathcal{F}} \equiv \braket{\Psi | \mathcal{F} | \Psi}$ and $\sqrbr{\mathcal{F}} \equiv \sqrt{\braket{\mathcal{F}^2} - \braket{\mathcal{F}}^2}$, respectively. 
$\braket{\mathcal{F}}_\star$ and $\sqrbr{\mathcal{F}}_\star$ should be evaluated for $\ket{\Psi_\star}$ instead  of $\ket{\Psi}$. 
Then the following relation holds \cite{Hasegawa:2021:QTURLEPRL}:
\begin{equation}
\left(\frac{\sqrbr{\mathcal{F}}+\sqrbr{\mathcal{F}}_{\star}}{\braket{\mathcal{F}}-\braket{\mathcal{F}}_{\star}}\right)^{2}\ge\frac{1}{\eta^{-1}-1},
\label{eq:main_result_prl}
\end{equation}where $\eta = |\braket{\Psi_\star|\Psi}|^2$.
Equation~\eqref{eq:main_result_prl} is a tighter version of the inequality derived in Ref.~\cite{Holevo:1973:CRI}. We obtained a quantum TUR for quantum Markov chains that end at a fixed time using Eq. ~\eqref{eq:main_result_prl}, where the observable is a counting observable and the lower bound is defined by the Loschmidt echo. The results obtained in Ref.~\cite{Hasegawa:2021:QTURLEPRL} can be regarded as a quantum analog of TUR, because the entropy production in stochastic thermodynamics also characterizes the extent of irreversibility due to the time-reversal operation. 

\begin{figure}
\includegraphics[width=8cm]{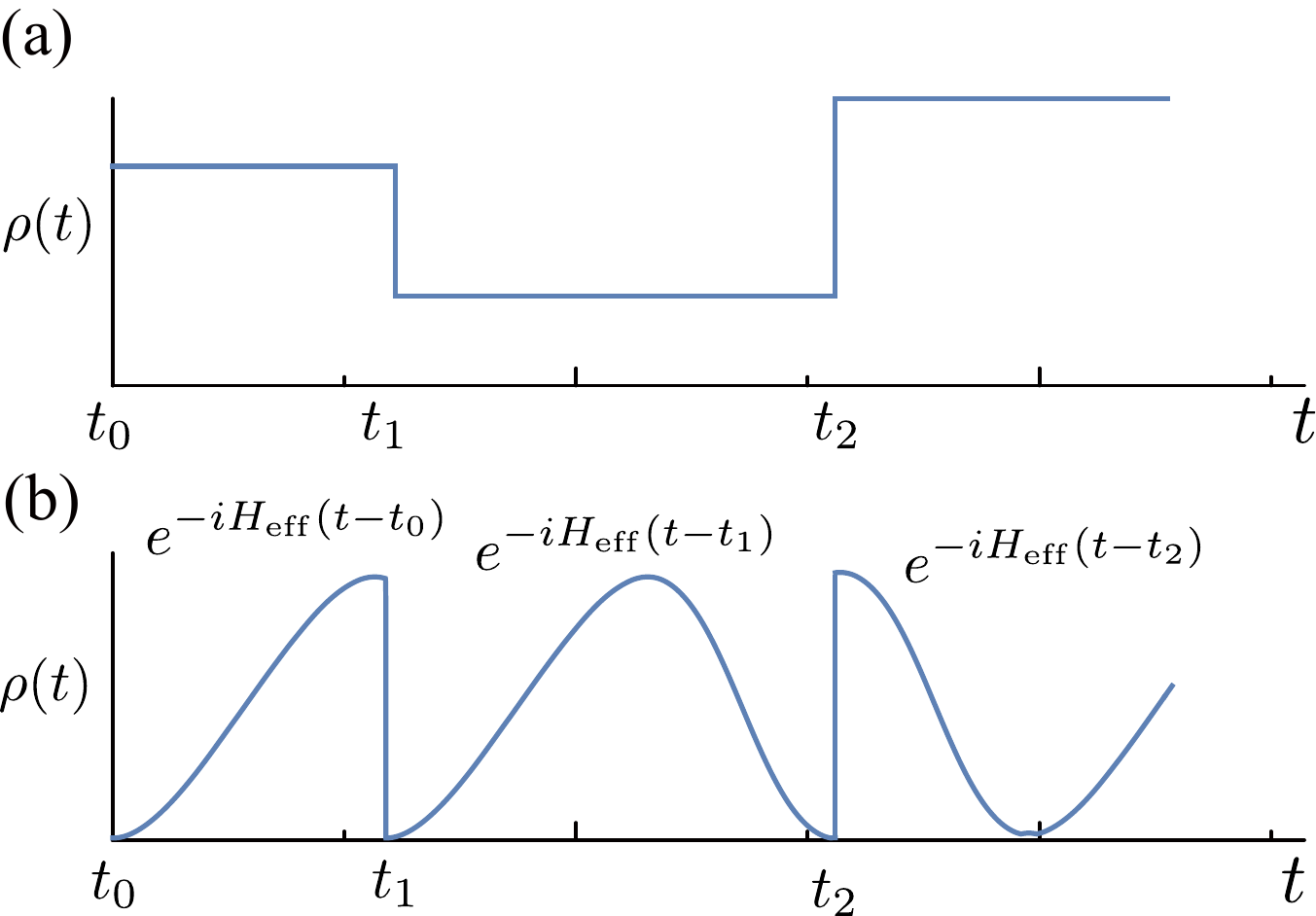} 
\caption{
Comparison of trajectories in (a) classical and (b) quantum Markov chains. 
In a quantum Markov chain,
between the $i$th and $i+1$th jump events,
the state can change via $e^{-iH_\mathrm{eff}(t-t_i)}$,
where $t_i$ is time stamp of the $i$th jump.
On the other hand, the state remains unchanged between consecutive jump events for a classical Markov chain.
\label{fig:classical_vs_quantum}}
\end{figure}

Now, we derive a TUR for first passage processes in quantum Markov chains using Eq.~\eqref{eq:main_result_prl}.
In particular, we consider quantum Markov chains that stop after $K$ jump events, where $K \in \{1, 2, 3,\ldots\}$. 
Let $\rho_S(t)$ be a density operator of the principal system $S$ at time $t$.
We assume that the dynamics of $\rho_S(t)$ is governed by the Lindblad equation:
\begin{equation}
\dot{\rho}_S = \mathcal{L}\rho_S \equiv  -i\left[H_S,\rho_S\right]+\sum_{m=1}^{M}\mathcal{D}(\rho_S,L_{m}),
\label{eq:Lindblad_PRE_def}
\end{equation}where $\mathcal{L}$ is the Lindblad super-operator, $H_S$ is a system Hamiltonian, $L_m$ represents jump operators, $M$ is the number of jump operators, and $\mathcal{D}$ is the dissipator defined by
\begin{equation}
\mathcal{D}(\rho_{S},L)\equiv L\rho_{S}L^{\dagger}-\frac{\left\{ L^{\dagger}L,\rho_{S}\right\} }{2}.
\label{eq:dissipator_def}
\end{equation}Here, $\{\bullet,\bullet\}$ denotes the anti-commutator. 
The solution of Eq.~\eqref{eq:Lindblad_PRE_def} corresponds to the dynamics when we do not measure the environment $E$. However, when we measure the environment $E$, the system exhibits stochastic dynamics depending on the measurement record. 
Such stochastic dynamics conditioned on the measurement record is referred to as a quantum trajectory.
Figures~\ref{fig:classical_vs_quantum}(a) and (b) show trajectories of
classical and quantum Markov chains, respectively.
For the classical trajectory, the state only varies via
stochastic jumps, whereas the system undergoes smooth continuous evolution and sudden discontinuous jumps for the quantum trajectory. 
We can represent Eq.~\eqref{eq:Lindblad_PRE_def} by the following alternative expression:
\begin{equation}
    \dot{\rho}_{S}=-i(H_{\mathrm{eff}}\rho_{S}-\rho_{S}H_{\mathrm{eff}}^{\dagger})+\sum_{m=1}^{M}L_{m}\rho_{S}L_{m}^{\dagger},
    \label{eq:Lindblad_alt_def}
\end{equation}
where $H_\mathrm{eff}$ is an effective Hamiltonian:
\begin{equation}
H_{\mathrm{eff}}\equiv H_{S}-\frac{i}{2}\sum_{m=1}^{M}L_{m}^{\dagger}L_{m}.
\label{eq:Heff_def}
\end{equation}
Equation~\eqref{eq:Lindblad_alt_def} shows that
the dynamics of $\rho_S$ is governed by two contributions:
the effective Hamiltonian term $-i(H_{\mathrm{eff}}\rho_{S}-\rho_{S}H_{\mathrm{eff}}^{\dagger})$,
which induces a continuous smooth state change, and
the jump operator term $\sum_{m=1}^{M}L_{m}\rho_{S}L_{m}^{\dagger}$,
which causes a discontinuous jump event. 
Suppose that the state of $S$ right after a jump event is $\ket{\psi_S}$. Continuous evolution in the quantum trajectory is given by $e^{-iH_\mathrm{eff}t} \ket{\psi_S}$, which is not normalized, and the next jump is induced by the $m$th jump operator $L_m$ after a waiting time $w$.
Therefore, from the state after the first jump event to the state after the second jump event, the state is transformed as
\begin{equation}
\ket{\psi_{S}}\to\frac{Y(w,m)\ket{\psi_{S}}}{\sqrt{\braket{\psi_{S}|Y^{\dagger}(w,m)Y(w,m)|\psi_{S}}}},
\label{eq:one_step_evolution}
\end{equation}where $Y(w,m) \equiv L_m e^{-iH_\mathrm{eff}w}$. 
$Y(w,m)$ satisfies a completeness relation $\sum_{m=1}^{M}\int_{0}^{\infty}dw\,Y^{\dagger}(w,m)Y(w,m)=\mathbb{I}_{S}$, which can be shown by 
\begin{align}
&\sum_{m=1}^{M}\int_{0}^{\infty}dw\,e^{iwH_{\mathrm{eff}}^{\dagger}}L_{m}^{\dagger}L_{m}e^{-iwH_{\mathrm{eff}}}\nonumber\\&=-\int_{0}^{\infty}dw\,\frac{d}{dw}\left[e^{iwH_{\mathrm{eff}}^{\dagger}}e^{-iwH_{\mathrm{eff}}}\right]\nonumber\\&=\mathbb{I}_{S}-\left.e^{iwH_{\mathrm{eff}}^{\dagger}}e^{-iwH_{\mathrm{eff}}}\right|_{w\to\infty}
\label{eq:norm_cond}
\end{align}Therefore, the Kraus representation describing the evolution 
from the first to the second jump events
is given by
\begin{equation}
\mathcal{Z}(\bullet)\equiv\sum_{m=1}^{M}\int_{0}^{\infty}dw\,Y(w,m)\bullet Y^{\dagger}(w,m).
\label{eq:Z_channel_d}
\end{equation}

We employ the input-output formalism for the first passage process with a fixed number of jumps as in Ref.~\cite{Kiukas:2015:MPS}. 
The input-output formalism is also known as
the continuous matrix product state. 
If we replace $V_m$ in Eq.~\eqref{eq:Kraus_repr} by $Y(w,m)$, the Kraus representation of Eq.~\eqref{eq:Z_channel_d} can be represented as an interaction between the principal system $S$ and the environment $E$ as follows:
\begin{equation}
    \ket{\Psi}=\sum_{m=1}^{M}\int_{0}^{\infty}dw\,Y(w,m)\ket{\psi_{S}}\otimes\ket{w,m},
    \label{eq:one_step_def}
\end{equation}where $\ket{w,m}$ constitutes an orthonormal basis in $E$ as $\braket{w',m'|w,m}=\delta(w'-w)\delta_{m'm}$.

\begin{figure}
\includegraphics[width=8.5cm]{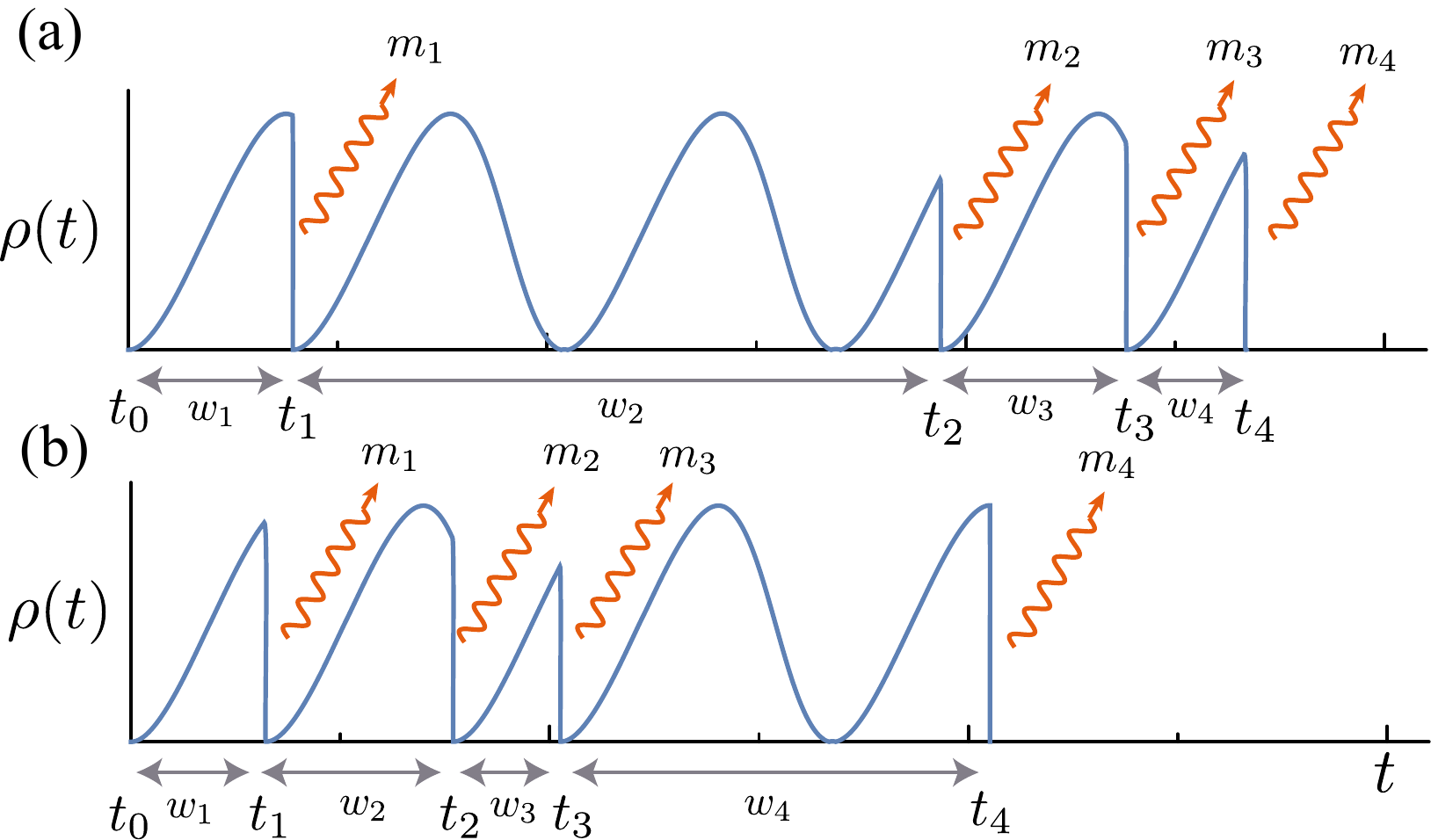} \caption{
Quantum Markov chain that stops after a fixed number of jumps ($K=4$).
(a) and (b) are different realizations of the process, both stopping when they undergo $K = 4$ jump events. 
$m_i$, $t_i$, and $w_i$ denote the $i$th output, time stamp of the $i$th output, and waiting time between the $(i-1)$th and $i$th jump events, respectively. 
\label{fig:ponch_PRE}}
\end{figure}

We consider a quantum Markov chain that stops after $K$ jump events. 
Let $\ket{\Psi_K}$ be a composite state in $S+E$ after $K$ jump events. 
Let $m_i$ be the output of the $i$th jump, $t_i$ be the time stamp of the $i$th jump ($t_0 = 0$), and $w_i$ be the waiting time between the $i-1$th and $i$th jump events (i.e., $w_i \equiv t_i - t_{i-1}$).
Figures~\ref{fig:ponch_PRE}(a) and (b) are examples of the process, where the trajectories stop when they undergo $K=4$ jump events.
We observe that the first passage time is a random variable, and it is different for these two cases. 
Repeating Eq.~\eqref{eq:one_step_def} $K$ times, $\ket{\Psi_K}$ is expressed by
\begin{widetext}
\begin{equation}
\ket{\Psi_{K}}=\sum_{m_{K},\cdots,m_{1}}\int_{0}^{\infty}d\boldsymbol{w}\,Y(w_{K},m_{K})\cdots Y(w_{1},m_{1})\ket{\psi_{S}}\otimes\ket{(w_{K},m_{K}),\cdots,(w_{1},m_{1})},
\label{eq:MPS_PRE_def}
\end{equation}\end{widetext}where $\int_0^\infty d\boldsymbol{w}$ is an abbreviation for $\int_0^\infty dw_1 \int_0^\infty dw_2 \cdots \int_0^\infty dw_K$. We used Eq.~\inputUoutput{} in Ref.~\cite{Hasegawa:2021:QTURLEPRL} for expressing the input-output state for constant-time quantum Markov chains. Equation~\eqref{eq:MPS_PRE_def} shows a constant jump case of Eq.~\inputUoutput{} in Ref.~\cite{Hasegawa:2021:QTURLEPRL}. 

When we perform a continuous measurement, we obtain records of jump events $\boldsymbol{m}\equiv[m_1,m_2,\ldots,m_K]$ and their time stamp  $\boldsymbol{t}\equiv[t_1,t_2,\ldots,t_K]$, as shown in Fig.~\ref{fig:ponch_PRE}. 
Because knowing the waiting time $\boldsymbol{w}\equiv[w_1,w_2,\ldots,w_K]$ is equivalent to the time stamp $\boldsymbol{t}$, $\boldsymbol{m}$, and $\boldsymbol{t}$ can be obtained by applying the projector $\ket{\boldsymbol{w},\boldsymbol{m}}\bra{\boldsymbol{w},\boldsymbol{m}}$, where $\ket{\boldsymbol{w},\boldsymbol{m}} \equiv \ket{(w_K,m_K)}\otimes \cdots \otimes \ket{(w_1,m_1)}$, to $\ket{\Psi_K}$ in Eq.~\eqref{eq:MPS_PRE_def}. 
We consider the following Hermitian observable on $E$. 
\begin{equation}
\mathcal{O}\equiv \sum_{\boldsymbol{m}}\int_{0}^{\infty}d\boldsymbol{w}\,h(\boldsymbol{w},\boldsymbol{m})\ket{\boldsymbol{w},\boldsymbol{m}}\bra{\boldsymbol{w},\boldsymbol{m}},
\label{eq:observable_def}
\end{equation}where $h(\boldsymbol{w},\boldsymbol{m})$ is an arbitrary real
function of $\boldsymbol{w}$ and $\boldsymbol{m}$.

The Loschmidt echo considers the fidelity between the original $\ket{\Psi_K}$ and the perturbed $\ket{\Psi_{\star,K}}$ composite states in $S+E$, as shown in
Fig.~\ref{fig:LoschmidtEcho_PRE}(b). 
Let $H_{\star,S}$ and $L_{\star,m}$ be the perturbed Hamiltonian and jump operators, respectively. 
The perturbed state $\ket{\Psi_{\star,K}}$ is expressed by Eq.~\eqref{eq:MPS_PRE_def}, where $Y(w,m)$ should be
replaced by the perturbed operator $Y_{\star}(w,m)\equiv L_{\star,m}e^{-iH_{\star,\mathrm{eff}}w}$ with $H_{\star,\mathrm{eff}}\equiv H_{\star,S}-\frac{i}{2}\sum_{m=1}^{M}L_{\star,m}^{\dagger}L_{\star,m}$. 
Then the Loschmidt echo $\eta $ becomes 
\begin{equation}
\eta=\left|\mathrm{Tr}\left[\mathcal{Z}_{\star}^{K}(\rho_{S}(0))\right]\right|^{2},
\label{eq:Loschmidt_echo_K}
\end{equation}where $\rho_S(0)$ is the initial state in $S$ and $\mathcal{Z}_\star$ is a mapping defined by
\begin{equation}
\mathcal{Z}_{\star}(\bullet)\equiv\sum_{m}\int_{0}^{\infty}dw\,Y(w,m)\bullet Y_{\star}^{\dagger}(w,m).
\label{eq:Z_star_def}
\end{equation}Here, $\mathcal{A}^K$ denotes exponentiation of a mapping $\mathcal{A}$, i.e., 
\begin{equation}
\mathcal{A}^K \equiv \underbrace{\mathcal{A}\circ \mathcal{A}\circ \cdots \circ \mathcal{A}}_{K},
\label{eq:AK_def}
\end{equation}where $\circ$ denotes a composition. Note that the mapping $\mathcal{Z}_\star$ is not a Kraus representation, because $\mathrm{Tr}[\mathcal{Z}_\star(\rho_S)] \ne 1$ in general. $\mathcal{Z}_\star$ is the fixed jump number analog of a two-sided Lindblad operator \cite{Gammelmark:2014:QCRB,Molmer:2015:HypoTest}, which is used to calculate the Loschmidt echo in the fixed end time system. 
Composition of the mapping can be reduced to a matrix multiplication when we convert 
$\mathcal{Z}_\star$ from the Hilbert space to the Liouville space (Appendix~\ref{sec:liouville_representation}), thereby making it easy to calculate $\eta$. 

Substituting $\mathcal{F} = \mathbb{I}_S \otimes \mathcal{O}$ in Eq.~\eqref{eq:observable_def} and $\eta$ in Eq.~\eqref{eq:Loschmidt_echo_K}
into Eq.~\eqref{eq:main_result_prl}, we obtain
\begin{equation}
\left(\frac{\sqrbr{\mathcal{O}}+\sqrbr{\mathcal{O}}_{\star}}{\braket{\mathcal{O}}-\mathcal{\braket{\mathcal{O}}_{\star}}}\right)^{2}\ge\frac{1}{\left|\mathrm{Tr}\left[\mathcal{Z}_{\star}^{K}(\rho_{S}(0))\right]\right|^{-2}-1},
\label{eq:main_result_K}
\end{equation}which is the main result in this manuscript.
Equation~\eqref{eq:main_result_K} is a TUR for first passage processes in quantum Markov chains. 
The above calculations assume an initially pure state $\rho_S(0) = \ket{\psi_S}\bra{\psi_S}$,
but it is straightforward to generalize to an
initially mixed state through purification as was done in Ref.~\cite{Hasegawa:2021:QTURLEPRL}.
It is important to note that the Loschmidt echo can be computed using information of the principal system only, i.e., 
$H_S$, $H_{\star,S}$, $L_m$, and $L_{\star,m}$. 
Moreover, by using an ancilla qubit,
we can show that the Loschmidt echo of Eq.~\eqref{eq:Loschmidt_echo_K}
becomes a measurable quantity of a physical process (Appendix~\ref{sec:measurement}).

Equation~\eqref{eq:main_result_K} holds for arbitrary perturbed dynamics specified by $H_{\star,S}$ and $L_{\star,m}$. 
Next, we consider a specific perturbed dynamics where Eq.~\eqref{eq:main_result_K} reduces to a more intuitive form. 
Let us consider the following perturbed operators:
\begin{align}
L_{\star,m}&=\sqrt{1+\varepsilon}L_{m},\label{eq:L_star_alpha_def}\\
H_{\star,S}&=(1+\varepsilon)H_{S},\label{eq:HS_star_alpha_def}
\end{align}where $\varepsilon$ is a real value parameter. We take the $\varepsilon \to 0$ limit. 
Because the Lindblad equation for the perturbed dynamics specified by Eqs.~\eqref{eq:L_star_alpha_def} and \eqref{eq:HS_star_alpha_def} is given by $\dot{\rho} = (1+\varepsilon) \mathcal{L}(\rho_S)$, 
the perturbed dynamics is the same as the original dynamics, except for the time scale. 
We may define a first passage time observable $\mathcal{O}_f$, which is a subset of $\mathcal{O}$, where the subscript $f$ denotes the capital of the first passage time. 
The time scale of the perturbed dynamics is $(1+\varepsilon)$ times faster than that of the original dynamics.
We assume that the first passage time observable $\mathcal{O}_f$ scales as
\begin{align}
    \braket{\mathcal{O}_{f}}_{\star}&=\frac{1}{1+\varepsilon}\braket{\mathcal{O}_{f}},\label{eq:Of_mean_scale}\\
    \sqrbr{\mathcal{O}_{f}}_{\star}&=\frac{1}{1+\varepsilon}\sqrbr{\mathcal{O}_{f}}.\label{eq:Of_var_scale}
\end{align}
For instance, when $h(\boldsymbol{w},\boldsymbol{m}) = \sum_{i=1}^{K} w_i = t_K$ in Eq.~\eqref{eq:observable_def}, the observable corresponds to the first passage time to reach $K$ jump events. 
Apparently, this first passage time satisfies Eqs.~\eqref{eq:Of_mean_scale} and \eqref{eq:Of_var_scale}.
With this scaling relation, the left hand side of Eq.~\eqref{eq:main_result_K} becomes
\begin{equation}
\frac{\sqrbr{\mathcal{O}_{f}}+\sqrbr{\mathcal{O}_{f}}_{\star}}{\braket{\mathcal{O}_{f}}-\braket{\mathcal{O}_{f}}_{\star}}=\frac{\varepsilon+2}{\varepsilon}\frac{\sqrbr{\mathcal{O}_{f}}}{\braket{\mathcal{O}_{f}}}.
\label{eq:observable_scaling}
\end{equation}
Next, we evaluate the right-hand side of Eq.~\eqref{eq:main_result_K}.
When we employ the perturbed Hamiltonian and jump operators in Eqs.~\eqref{eq:HS_star_alpha_def} and
\eqref{eq:L_star_alpha_def}, respectively, the $\varepsilon$ dependence of the composite state of the perturbed dynamics can be expressed as
\begin{equation}
\ket{\Psi_{\star,K}} = \ket{\Psi_K(\varepsilon)}.
\label{eq:Psi_K_alpha_def}
\end{equation}When $\varepsilon=0$, Eq.~\eqref{eq:Psi_K_alpha_def} reduces to the unperturbed state, $\ket{\Psi_K(\varepsilon = 0)} = \ket{\Psi_K}$. 
Let us consider the quantum Fisher information \cite{Paris:2009:QFI,Liu:2019:QFisherReviewJPA}, which is obtained by the following calculation:
\begin{align}
\mathcal{J}_K(\alpha)&=4\braket{\partial_{\alpha}\Psi_{K}(\alpha)|\partial_{\alpha}\Psi_{K}(\alpha)}\nonumber\\&+4(\braket{\partial_{\alpha}\Psi_{K}(\alpha)|\Psi_{K}(\alpha)})^{2},
\label{eq:FQ_def}
\end{align}where $\ket{\partial_{\alpha}\Psi_{K}(\alpha)}\equiv(d/d\alpha)\ket{\Psi_{K}(\alpha)}$. It is known that the fidelity and quantum Fisher information are related via \cite{Braunstein:1994:QFI}
\begin{equation}
\mathcal{J}_K(\alpha)=\frac{8}{\varepsilon^{2}}\left[1-\left|\braket{\Psi_{K}(\alpha+\varepsilon)|\Psi_{K}(\alpha)}\right|\right],
\label{eq:Fisher_and_fidelity}
\end{equation}where we consider the $\varepsilon \to 0$ limit. Substituting Eqs.~\eqref{eq:observable_scaling} and \eqref{eq:Fisher_and_fidelity} into Eq.~\eqref{eq:main_result_K}, 
we obtain
\begin{equation}
\frac{\sqrbr{\mathcal{O}_{f}}^{2}}{\braket{\mathcal{O}_{f}}^{2}}\ge\frac{1}{\mathcal{J}_{K}(0)}.
\label{eq:alpha_main_result}
\end{equation}Equation~\eqref{eq:alpha_main_result} provides a TUR for the first passage time in quantum Markov chains. 
As will be shown later, when we only consider classical stochastic processes, we obtain $\mathcal{J}_K(0) \to K$. 
The quantum Fisher information gives the fundamental limit of the quantum parameter estimation \cite{Helstrom:1976:QuantumEst,Hotta:2004:QEstimation,Paris:2009:QFI,Liu:2019:QFisherReviewJPA}. 
Moreover, it plays a central role in the quantum speed limit \cite{Deffner:2017:QSLReview,Frowis:2012:FisherQSL,Taddei:2013:QSL} and a quantum TUR \cite{Hasegawa:2020:QTURPRL}. 
If we measure quantum systems and obtain a measurement output,
the obtained output can be treated classically, and we observe that the quantum nature does not come into play. 
Therefore, we let $\mathcal{I}_K(\alpha;\mathcal{M}_{SE})$ be the classical Fisher information, which is obtained by applying a positive operator-valued measure (POVM) $\mathcal{M}_{SE}$ to $\ket{\Psi_K(\alpha)}$. 
Because a record of continuous measurement can be obtained by measuring $\ket{\Psi_K(\alpha)}$
by the projector $\ket{\boldsymbol{w},\boldsymbol{m}}\bra{\boldsymbol{w},\boldsymbol{m}}$, $\mathcal{I}_K^\mathrm{cm}(\alpha)$ is represented as
\begin{equation}
\mathcal{I}_{K}^{\mathrm{cm}}(\alpha)=\mathcal{I}_{K}(\alpha;\mathbb{I}_{S}\otimes\{\ket{\boldsymbol{w},\boldsymbol{m}}\bra{\boldsymbol{w},\boldsymbol{m}}\}).
\label{eq:IC_cm_def}
\end{equation}Because the quantum Fisher information is larger than the classical counterpart, we have
\begin{equation}
\mathcal{I}_{K}^{\mathrm{cm}}(\alpha)\le\mathcal{J}_{K}(\alpha).
\label{eq:IC_le_IQ}
\end{equation}
We have derived Eq.~\eqref{eq:alpha_main_result} using the lower bound of the Hellinger distance and 
considering a particular perturbed dynamics specified by Eqs.~\eqref{eq:L_star_alpha_def} and \eqref{eq:HS_star_alpha_def}.
As shown in Appendix~\ref{sec:QCR_derivation}, Eq.~\eqref{eq:alpha_main_result} can also be derived through the classical Cram\'er--Rao inequality,
which was employed to derive classical TURs \cite{Hasegawa:2019:CRI,Dechant:2019:MTUR}.

\begin{figure}
\includegraphics[width=8cm]{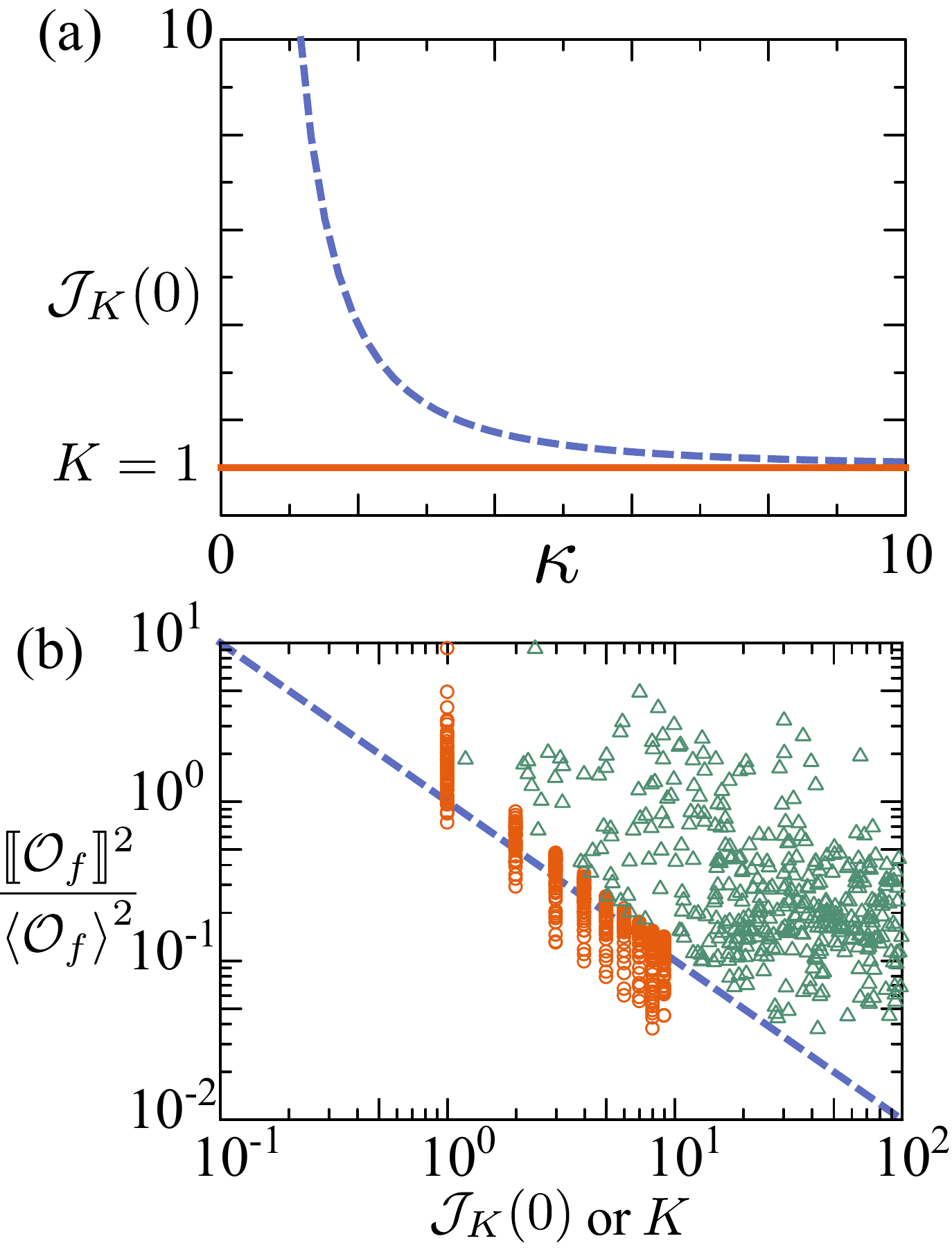} 
\caption{
(a) Quantum Fisher information $\mathcal{J}_K(0)$ as a function of $\kappa$. The dashed line denotes $\mathcal{J}_K(0)$ and the solid line denotes $K$, which is the classical limit of $\mathcal{J}_K(0)$. The parameters are $\Delta = \Omega = 1.0$ and $K=1$. 
(b) Precision $\sqrbr{\mathcal{O}_{f}}^{2}/\braket{\mathcal{O}_{f}}^{2}$ as a function of the quantum Fisher information $\mathcal{J}_K(0)$ (triangles) and the jump number $K$ (circles) for random realizations. The dashed line corresponds to $1/\mathcal{J}_K(0)$ for the triangles and $1/K$ for the circles. The parameters are randomly selected from $\Delta \in [0.1, 3.0]$, $\Omega \in [0.1, 3.0]$, and $\kappa \in [1.0, 3.0]$. 
\label{fig:sim_result}}
\end{figure}

Next, we obtain the classical limit of Eq.~\eqref{eq:alpha_main_result}. 
We consider a classical Markov chain with $N_S$ states, $B_1,B_2,...,B_{N_S}$, and 
take $\gamma_{ji}$ to be a time-independent transition rate from $B_i$ to $B_j$. 
Eq.~\eqref{eq:alpha_main_result} then reduces to the following simple relation:
\begin{equation}
\frac{\sqrbr{\mathcal{O}_f}^{2}}{\braket{\mathcal{O}_f}^{2}}\ge\frac{1}{K},
\label{eq:classical_FPTTUR}
\end{equation} as shown in Appendix~\ref{sec:classical_limit}.
Equation~\eqref{eq:classical_FPTTUR} corresponds to a specific case of the result reported in Ref.~\cite{Hiura:2021:KUR}.
Equation~\eqref{eq:classical_FPTTUR} shows that for first passage processes that stop after
$K$ jumps, the lower bound does not depend on the details of the dynamics or the initial distribution.

\section{Example\label{sec:example}}
We apply the main result, Eq.~\eqref{eq:main_result_K}, to a two-level atom driven by a classical laser field. 
Let $\ket{\epsilon_e}$ and $\ket{\epsilon_g}$ be the excited and ground states, respectively. 
The Hamiltonian $H_S$ and a jump operator $L$ are given by
\begin{align}
H_{S}&=\Delta\ket{\epsilon_{e}}\bra{\epsilon_{e}}+\frac{\Omega}{2}\left[\ket{\epsilon_{e}}\bra{\epsilon_{g}}+\ket{\epsilon_{g}}\bra{\epsilon_{e}}\right],\label{eq:HS_atom_def}\\
L&=\sqrt{\kappa}\ket{\epsilon_{g}}\bra{\epsilon_{e}},\label{eq:L_atom_def}
\end{align}where $\Delta$ is the detuning between the laser field and the atomic-transition frequencies, $\Omega$ is the Rabi-oscillation frequency, and $\kappa$ is the decay rate. $L$ induces a jump from $\ket{\epsilon_e}$ to $\ket{\epsilon_g}$. 
We first calculate the parameter dependence of the quantum Fisher information $\mathcal{J}_K(0)$ in Eq.~\eqref{eq:alpha_main_result} on the decay rate $\kappa$. 
Figure~\ref{fig:sim_result}(a) shows $\mathcal{J}_K(0)$ as a function of $\kappa$, with the parameters given in the caption. The dashed line denotes $\mathcal{J}_K(0)$ and the solid line shows $K$, which is the classical limit of the lower bound, as shown in Eq.~\eqref{eq:classical_FPTTUR}. 
We see that $\mathcal{J}_K(0)$ reduces to $K$ as $\kappa\to \infty$. 
Because jumps in the Lindblad dynamics become dominant compared to the dynamics induced by $e^{-iH_\mathrm{eff}t}$ for $\kappa \to \infty$, the dynamics becomes closed to a classical Markov chain. 
Moreover, we observe that $\mathcal{J}_K(0)$ is always larger than $K$, which indicates that the lower bound of the TUR is smaller for this quantum Markov chain. 

To confirm that the fluctuation of the first passage observable $\mathcal{O}_f$ is smaller for the quantum Markov chain, we perform a computer simulation. 
We randomly generate $\Delta$, $\Omega$, and $\kappa$ and calculate $\sqrbr{\mathcal{O}_{f}}^{2}/\braket{\mathcal{O}_{f}}^{2}$.
Specifically, we consider $\mathcal{O}_f$ with $h(\boldsymbol{w},\boldsymbol{m}) = \sum_{i=1}^K w_i =t_K$, which gives the first passage time needed to undergo $K$ jump events.
We plot $\sqrbr{\mathcal{O}_{f}}^{2}/\braket{\mathcal{O}_{f}}^{2}$ as a function of the quantum Fisher information $\mathcal{J}_K(0)$ with triangles, where the dashed line denotes the lower bound, as shown in Fig.~\ref{fig:sim_result}(b). We confirm that all realizations are above the dashed line, which numerically verifies Eq.~\eqref{eq:alpha_main_result}.
We have shown that Eq.~\eqref{eq:alpha_main_result} reduces to Eq.~\eqref{eq:classical_FPTTUR} in a classical case, where the lower bound is given by $1/K$. Therefore, we also check whether $\sqrbr{\mathcal{O}_{f}}^{2}/\braket{\mathcal{O}_{f}}^{2}$ can be bounded from below by $1/K$. We plot $\sqrbr{\mathcal{O}_{f}}^{2}/\braket{\mathcal{O}_{f}}^{2}$ as a function of $K$ with circles, where the dashed line now describes $1/K$, as shown in Fig.~\ref{fig:sim_result}(b). 
We observe that some circles are below the dashed line, which indicates that the observable $\mathcal{O}_f$ cannot be bounded from below by $1/K$.
As explained in Fig.~\ref{fig:sim_result}(a),
this enhancement of precision can be ascribed to
the fact that $\mathcal{J}_K(0)$ is larger than
$K$ for smaller $\kappa$,
where the state change in the dynamics 
is mostly induced by $H_\mathrm{eff}$. 
Therefore, the precision enhancement in the quantum Markov chain
is due to the state change via the effective Hamiltonian. 
Quantum-induced precision enhancement has been reported for quantum Markov chains with a fixed end time \cite{Carollo:2019:QuantumLDP,Hasegawa:2020:QTURPRL}. Recently, it was reported that quantum coherence can improve the precision of a quantum heat engine \cite{Kalaee:2021:QTUR}.

\section{Conclusion}

In this manuscript, we derived a TUR for a first passage time in quantum Markov chains. Furthermore, we used a derivation technique developed in Ref.~\cite{Hasegawa:2021:QTURLEPRL}, where the fluctuation of observables is bounded from below by the Loschmidt echo. Our approach is quite general, and it can be applied to many quantum systems, which could not  be handled with the previous derivations. 
A possible application of our approach is an extension to non-Markovian systems such as time-delayed systems \cite{Grimsmo:2015:TimeDelay}. 
Moreover, recently, the notion of quantum thermodynamics 
has been generalized to
quantum field theory \cite{Bartolotta:2018:JEQFT,Ortega:2019:WorkQFT}. Because the input-output formalism itself is a quantum field, 
we expect that uncertainty relations for quantum field theory can be derived using our technique.

\begin{acknowledgments}
This work was supported by the Ministry of Education, Culture, Sports, Science and Technology (MEXT) KAKENHI Grant No.~JP19K12153.
\end{acknowledgments}

\appendix

\section{Liouville space representation\label{sec:liouville_representation}}
The mapping $\mathcal{Z}_\star$ in the Hilbert space is shown in Eq.~\eqref{eq:Z_star_def}. 
Because the calculation of $\mathcal{Z}_\star^K$ is a computationally expensive task, 
we can use the Liouville space representation of $\mathcal{Z}_\star$. 
The Liouville space representation is advantageous because the mapping can be realized via matrix multiplication. 
In the Hilbert space, the density operator is
\begin{equation}
\rho = \sum_{i, j} \varrho_{ij}\ket{i}\bra{j},
\label{eq:Hilbert_density_def}
\end{equation}where $\ket{i}$ is an orthonormal basis. It can be represented in the Liouville space as follows \cite{Landi:2018:TextBook}:
\begin{equation}
\mathrm{vec}(\rho)=\sum_{i,j}\varrho_{ij}\ket{j}\otimes\ket{i}.
\label{eq:Liouville_density_def}
\end{equation}Let $\mathsf{A}$, $\mathsf{B}$, and $\mathsf{C}$ be arbitrary matrices in the Hilbert space. Then the following relation holds:
\begin{equation}
\mathrm{vec}(\mathsf{A}\mathsf{B}\mathsf{C})=(\mathsf{C}^{\top}\otimes\mathsf{A})\mathrm{vec}(\mathsf{B}),
\label{eq:ABC_vec}
\end{equation}where $\top$ denotes the transpose. 
When the dimensions of the operators in the Hilbert space is $d\times d$, their corresponding Liouville space representations have
the dimensions of $d^2\times d^2$. 
Using Eq.~\eqref{eq:ABC_vec}, we obtain $\mathcal{Z}_\star$ [Eq.~\eqref{eq:Z_star_def}] in the Liouville space as follows:
\begin{equation}
\mathfrak{Z}_{\star}\equiv \sum_{m=1}^{M}\int_{0}^{\infty}dw\,Y_{\star}^{*}(w,m)\otimes Y(w,m),
\label{eq:Z_map_Liouville_def}
\end{equation}where the superscript $*$ denotes complex conjugate. Then, exponentiation of a mapping, $\mathcal{Z}_\star^K (\rho_S)$, is simply realized by a matrix power $\mathfrak{Z}_{\star}^{K}\mathrm{vec}(\rho_{S})$. 

\section{Measurement of Loschmidt echo\label{sec:measurement}}

We here show a method for obtaining the
Loschmidt echo given by Eq.~\eqref{eq:Loschmidt_echo_K} through 
measurement of a physical process. 

We first review the well known approach for calculating the Loschmidt echo for closed quantum dynamics. 
We can measure the Loschmidt echo for closed quantum dynamics by introducing an ancilla qubit.
Let $H$ and $H_\star$ be Hamiltonian operators 
of the original and the perturbed dynamics, respectively,
with which
we want to calculate the Loschmidt echo, and 
let $\ket{\mathfrak{0}}$ and $\ket{\mathfrak{1}}$ be the two states of the ancilla qubit. 
The Loschmidt echo for closed quantum dynamics is given by Eq.~\eqref{eq:Loschmidt_PRE_def}.
We define the Hamiltonian in the composite system comprising the ancilla and system as follows:
\begin{equation}
\tilde{H}\equiv\ket{\mathfrak{0}}\bra{\mathfrak{0}}\otimes H+\ket{\mathfrak{1}}\bra{\mathfrak{1}}\otimes H_{\star}=\left[\begin{array}{cc}
H & 0\\
0 & H_{\star}
\end{array}\right].
\label{eq:HSA_def}
\end{equation}
The initial state of the composite system is $\ket{\tilde{\Psi}(0)}=\frac{1}{\sqrt{2}}(\ket{\mathfrak{0}}+\ket{\mathfrak{1}})\otimes\ket{\Psi(0)}$, where $\ket{\Psi(0)}$ is the initial state of the original system. 
The state of the ancilla after the time evolution is given by
\begin{align}
    &\mathrm{Tr}_{S}\left[e^{-i\tilde{H}\tau}\ket{\tilde{\Psi}(0)}\bra{\tilde{\Psi}(0)}e^{i\tilde{H}\tau}\right]\nonumber\\&=\frac{1}{2}\left[\begin{array}{cc}
1 & \braket{\Psi_{\star}(\tau)|\Psi(\tau)}\\
\braket{\Psi(\tau)|\Psi_{\star}(\tau)} & 1
\end{array}\right],
\label{eq:ancilla_state_tau}
\end{align}where $\mathrm{Tr}_S$ is the trace operation
with respect to the principal system $S$, $\ket{\Psi(\tau)} = e^{-iH\tau}\ket{\Psi(0)}$,
and $\ket{\Psi_\star(\tau)} = e^{-i H_\star \tau}\ket{\Psi(0)}$.
From Eq.~\eqref{eq:ancilla_state_tau}, we see that
the Loschmidt echo $|\braket{\Psi_\star(\tau)|\Psi(\tau)}|$ can be
obtained by measuring the $\ket{\mathfrak{0}}\bra{\mathfrak{1}}$ (or $\ket{\mathfrak{1}}\bra{\mathfrak{0}}$) element of the ancilla.

The Loschmidt echo for a Lindblad equation
can be calculated in a similar way.
For a system that ends at a constant time, the Loschmidt echo can be calculated by
the procedure shown in Ref.~\cite{Molmer:2015:HypoTest}.
We here show how we can measure the Loschmidt echo for 
a constant jump case. 
Similar to Eq.~\eqref{eq:HSA_def}, we prepare an ancilla qubit and 
define the Hamiltonian and jump operators in the ancilla and system
as follows:
\begin{align}
    \tilde{H}_{S}&\equiv \ket{\mathfrak{0}}\bra{\mathfrak{0}}\otimes H_{S}+\ket{\mathfrak{1}}\bra{\mathfrak{1}}\otimes H_{\star,S}=\left[\begin{array}{cc}
H_{S} & 0\\
0 & H_{\star,S}
\end{array}\right],\label{eq:HS_tilde_def}\\
\tilde{L}_{m}&\equiv \ket{\mathfrak{0}}\bra{\mathfrak{0}}\otimes L_{m}+\ket{\mathfrak{1}}\bra{\mathfrak{1}}\otimes L_{\star,m}=\left[\begin{array}{cc}
L_{m} & 0\\
0 & L_{\star,m}
\end{array}\right].\label{eq:Lm_tilde_def}
\end{align}
We define the density operator in the ancilla and system as follows:
\begin{equation}
   \tilde{\rho}(t)=\frac{1}{2}\left[\begin{array}{cc}
\rho_{00}(t) & \rho_{01}(t)\\
\rho_{10}(t) & \rho_{11}(t)
\end{array}\right],
\label{eq:rhoSA_def}
\end{equation}where the initial state is set to be $\rho_{00}(0)=\rho_{01}(0)=\rho_{10}(0)=\rho_{11}(0)=\rho_S(0)$.
Similar to the closed quantum case, let $\tilde{\rho}(t)$ evolve
through $\tilde{H}_S$ [Eq.~\eqref{eq:HS_tilde_def}] and $\tilde{L}_m$ [Eq.~\eqref{eq:Lm_tilde_def}]. 
Specifically, suppose that $\tilde{\rho}(t)$ is governed by the following Lindblad equation:
\begin{equation}
   \dot{\tilde{\rho}}=-i\left[\tilde{H}_{S},\tilde{\rho}\right]+\sum_{m=1}^{M}\mathcal{D}(\tilde{\rho},\tilde{L}_{m}).
   \label{eq:Lindblad_SA_def}
\end{equation}
Let us consider a continuous measurement in Eq.~\eqref{eq:Lindblad_SA_def}. From Eq.~\eqref{eq:Z_channel_d}, the Kraus representation
describing the evolution from the first to the second jump events is
\begin{equation}
    \tilde{\mathcal{Z}}(\bullet)\equiv\sum_{m=1}^{M}\int_{0}^{\infty}dw\,\tilde{Y}(w,m)\bullet\tilde{Y}^{\dagger}(w,m),
    \label{eq:Kraus_SA_def}
\end{equation}where 
$\tilde{Y}(w,m)\equiv\tilde{L}_{m}e^{-i\tilde{H}_{\mathrm{eff}}w}$ with $\tilde{H}_{\mathrm{eff}}\equiv\tilde{H}_{S}-\frac{i}{2}\sum_{m=1}^{M}\tilde{L}_{m}^{\dagger}\tilde{L}_{m}$. 
Using Eqs.~\eqref{eq:HS_tilde_def} and \eqref{eq:Lm_tilde_def}, 
$\tilde{Y}(w,m)$ becomes
\begin{equation}
    \tilde{Y}(w,m)=\left[\begin{array}{cc}
Y(w,m) & 0\\
0 & Y_{\star}(w,m)
\end{array}\right].
\label{eq:Ytilde_calc}
\end{equation}Substituting 
Eqs.~\eqref{eq:rhoSA_def} and \eqref{eq:Ytilde_calc}
into Eq.~\eqref{eq:Kraus_SA_def} yields
\begin{widetext}
\begin{align}
    \tilde{\mathcal{Z}}(\tilde{\rho})&=\sum_{m=1}^{M}\int_{0}^{\infty}dw\,\frac{1}{2}\left[\begin{array}{cc}
Y(w,m)\rho_{00}Y^{\dagger}(w,m) & Y(w,m)\rho_{01}Y_{\star}^{\dagger}(w,m)\\
Y_{\star}(w,m)\rho_{01}Y^{\dagger}(w,m) & Y_{\star}(w,m)\rho_{11}Y_{\star}^{\dagger}(w,m)
\end{array}\right].\label{eq:Ztilde_calc}
\end{align}
Taking the trace with respect to the principal system in Eq.~\eqref{eq:Ztilde_calc}, we obtain the ancilla state after
a single jump event:
\begin{align}
\mathrm{Tr}_{S}\left[\tilde{\mathcal{Z}}(\tilde{\rho})\right]&=\sum_{m=1}^{M}\int_{0}^{\infty}dw\,\frac{1}{2}\left[\begin{array}{cc}
1 & \mathrm{Tr}\left[Y(w,m)\rho_{01}Y_{\star}^{\dagger}(w,m)\right]\\
\mathrm{Tr}\left[Y_{\star}(w,m)\rho_{01}Y^{\dagger}(w,m)\right] & 1
\end{array}\right].\label{eq:Ztilde_calc_trace}
\end{align}
\end{widetext}
Notably, the $\ket{\mathfrak{0}}\bra{\mathfrak{1}}$ element of the state of the ancilla 
is identical to $\mathcal{Z}_\star(\bullet)$ given in Eq.~\eqref{eq:Z_star_def} (when multiplied by $2$). 
Therefore, the Loschmidt echo for $K=1$ can be obtained by simply
measuring the $\ket{\mathfrak{0}}\bra{\mathfrak{1}}$ (or $\ket{\mathfrak{1}}\bra{\mathfrak{0}}$) element of the ancilla
after a single jump. 
Although the procedure explained above is for a single jump event, i.e., $K=1$, 
it is straightforward to generalize to $K>1$ cases. 
Therefore, we can measure the Loschmidt echo $\eta$ [Eq.~\eqref{eq:Loschmidt_echo_K}]
by first considering the ancilla qubit and 
performing a continuous measurement on the composite system 
comprising the ancilla and system.
After $K$ jump events, we measure the $\ket{\mathfrak{0}}\bra{\mathfrak{1}}$ (or $\ket{\mathfrak{1}}\bra{\mathfrak{0}}$)
element of the ancilla.
By repeating this procedure many times and 
calculating the average of the measurements, we obtain $\frac{1}{2}\mathrm{Tr}\left[\mathcal{Z}_{\star}^{K}(\rho_{S}(0))\right]$ which
directly gives the Loschmidt echo.

\begin{figure}
\includegraphics[width=8cm]{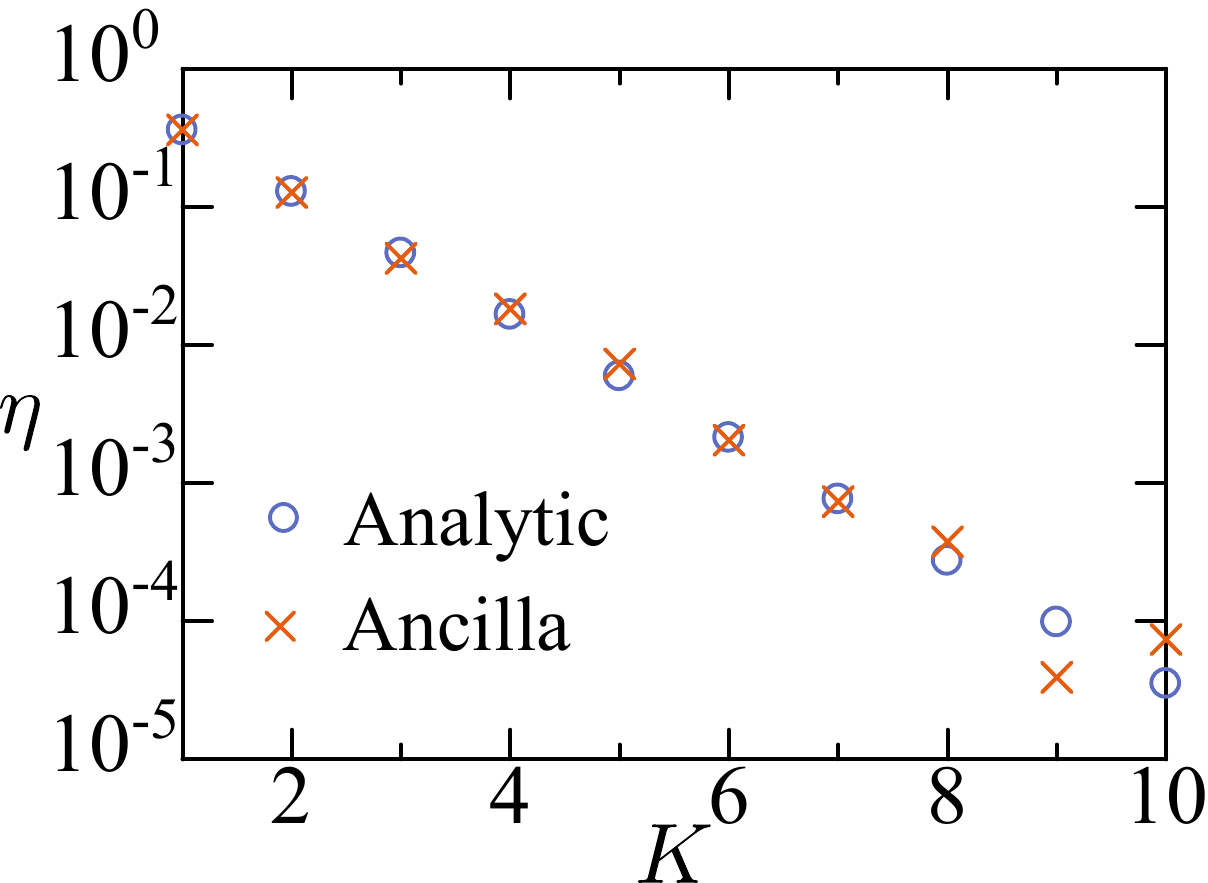} 
\caption{
Loschmidt echo $\eta$ as a function of
the number of jump events $K$, calculated by the analytic approach (circles) based on Eq.~\eqref{eq:Loschmidt_echo_K} and the ancilla approach (crosses) introduced in Sec.~\ref{sec:measurement}. 
The initial state is $\rho_S(0) = \ket{\epsilon_g}\bra{\epsilon_g}$. 
The parameters are $\Delta = 1.0$, $\Omega = 1.0$, and $\kappa = 2.0$ for the original process and $\Delta = 0.4$, $\Omega = 1.2$, and $\kappa = 0.5$ for the perturbed process. 
For the ancilla approach, 
the number of trials used to calculate each $\eta$ is $10000$.
\label{fig:ancilla_LE}}
\end{figure}

We numerically verify this approach of calculating the Loschmidt echo. 
We use a two-level atom driven by a classical laser field,
the same model as in Section~\ref{sec:example}.
Figure~\ref{fig:ancilla_LE} shows the Loschmidt echo $\eta$ as a function of $K$
obtained by the analytic and ancilla approaches, which
are shown by circles and crosses, respectively. 
The settings for the calculation are shown in the caption of 
Fig.~\ref{fig:ancilla_LE}. 
For the ancilla approach, $\eta$ is calculated by the
$\ket{\mathfrak{0}}\bra{\mathfrak{1}}$ element of the ancilla state after $K$ jump events. 
We can see that the ancilla approach yields very close
values for $K\le 7$, which numerically confirm the validity of
the ancilla approach. 
The disagreements for $K\ge 8$ are due to undersampling and hence can be improved by increasing the number of samples.

\section{Classical bound\label{sec:classical_limit}}
We derive the classical limit of Eq.~\eqref{eq:alpha_main_result}. 
We can represent $N_S$ states, $B_1,B_2,\ldots,B_{N_S}$,
in a classical Markov chain by
an orthonormal basis $\{\ket{b_1},\ket{b_2},\ldots,\ket{b_{N_S}}\}$.
To emulate classical Markov chains with quantum Markov chains, we consider the following Hamiltonian and jump operators in the Lindblad equation:
\begin{align}
H_{S}&=0,\label{eq:HS_classical}\\
L_{ji}&=\sqrt{\gamma_{ji}}\ket{b_{j}}\bra{b_{i}}\,\,\,\,\,(i\ne j).\label{eq:Lji_classical}
\end{align}Let $\varepsilon$ be a small perturbation parameter as defined in Eqs.~\eqref{eq:L_star_alpha_def} and \eqref{eq:HS_star_alpha_def}. The Hamiltonian and the jump operators for the perturbed dynamics become
\begin{align}
H_{\star,S}&=0,\label{eq:HS_star_classical}\\
L_{\star,ji}&=\sqrt{1+\varepsilon}\sqrt{\gamma_{ji}}\ket{b_{j}}\bra{b_{i}}\,\,\,\,\,(i\ne j).\label{eq:Lji_star_classical}
\end{align}We also consider the following initial density operator which emulates the classical probability distribution:
\begin{equation}
\rho_S(t) = \sum_i p_i(t) \ket{b_i}\bra{b_i},
\label{eq:classical_emulation_density}
\end{equation}where $p_i(t)$ is the classical probability distribution of being $\ket{b_i}$ at time $t$. 
Let $Y(w,i,j)\equiv L_{ji}e^{-iH_\mathrm{eff}w}$ (with $Y_\star(w,i,j)$ being defined in the same way). 
Substituting Eqs.~\eqref{eq:HS_classical}--\eqref{eq:classical_emulation_density} into Eq.~\eqref{eq:Z_star_def}, $\mathcal{Z}_\star$ in Eq.~\eqref{eq:Z_star_def} becomes
\begin{align}
\mathcal{Z}_{\star}(\rho_{S})&=\sum_{i\ne j}\int_{0}^{\infty}dwY(w,i,j)\rho_{S}Y_{\star}^{\dagger}(w,i,j)\nonumber\\&=\sum_{i\ne j}\int_{0}^{\infty}dw\,\sqrt{1+\varepsilon}\gamma_{ji}\nonumber\\&\times\exp\left[-\frac{w}{2}(2+\varepsilon)\sum_{k(\ne i)}\gamma_{ki}\right]p_{i}\ket{b_{j}}\bra{b_{j}}\nonumber\\&=\sum_{j}\sum_{i(\ne j)}\frac{2\sqrt{1+\varepsilon}}{(2+\varepsilon)}\frac{\gamma_{ji}}{\sum_{k(\ne i)}\gamma_{ki}}p_{i}\ket{b_{j}}\bra{b_{j}}\nonumber\\&=\sum_{j}\frac{2\sqrt{1+\varepsilon}}{(2+\varepsilon)}[\mathcal{B}\boldsymbol{p}]_{j}\ket{b_{j}}\bra{b_{j}}.\label{eq:A_classical}
\end{align}where $\boldsymbol{p} = [p_1,p_2,\ldots,p_N]^\top$ and $\mathcal{B}$ is a matrix defined by
\begin{equation}
\mathcal{B}_{ij} \equiv \begin{cases}
0 & i=j\\
\frac{\gamma_{ij}}{\sum_{k(\ne j)}\gamma_{kj}} & i\ne j
\end{cases}.
\label{eq:B_matrix_def}
\end{equation}Therefore we obtain
\begin{align}
\mathrm{Tr}\left[\mathcal{Z}_{\star}^{K}(\rho_{S})\right]&=\sum_{j}\left(\frac{2\sqrt{1+\varepsilon}}{2+\varepsilon}\right)^{K}\left[\mathcal{B}^{K}\boldsymbol{p}\right]_{j}.
\label{eq:TrS_A_classical}
\end{align}From Eq.~\eqref{eq:B_matrix_def}, $\mathcal{B}$ is a stochastic matrix $\sum_{i}\mathcal{B}_{ij}=1$, yielding 
\begin{equation}
\mathrm{Tr}\left[\mathcal{Z}_{\star}^{K}(\rho_{S})\right]=\left(\frac{2\sqrt{1+\varepsilon}}{2+\varepsilon}\right)^{K}.
\label{eq:TrS_AK}
\end{equation}$\mathcal{O}_f$ in the classical systems also satisfies the scaling condition of Eq.~\eqref{eq:observable_scaling}.
From Eqs.~\eqref{eq:observable_scaling} and \eqref{eq:TrS_AK}, Eq.~\eqref{eq:main_result_K} becomes
\begin{equation}
\frac{\sqrbr{\mathcal{O}_{f}}^{2}}{\braket{\mathcal{O}_{f}}^{2}}\ge\frac{1}{\left(\frac{\varepsilon+2}{\varepsilon}\right)^{2}\left[\left(\frac{2\sqrt{1+\varepsilon}}{2+\varepsilon}\right)^{-2K}-1\right]}\overset{\varepsilon\to0}{\to}\frac{1}{K},
\label{eq:classical_TUR_result}
\end{equation}where we used l'H{\^o}pital's rule for calculating the limit. 
Equation~\eqref{eq:classical_TUR_result} is a classical case of the main result [Eq.~\eqref{eq:classical_FPTTUR}].

\section{Derivation based on classical Cram\'er--Rao inequality\label{sec:QCR_derivation}}

Equation~\eqref{eq:alpha_main_result}, which is a particular case of the main result [Eq.~\eqref{eq:main_result_K}], can also be derived through the classical Cram\'er--Rao inequality. 
The classical Cram\'er--Rao inequality states
\begin{equation}
\frac{\sqrbr{\mathcal{O}}_{\alpha}^{2}}{\left(\partial_{\alpha}\braket{\mathcal{O}}_{\alpha}\right)^{2}}\ge\frac{1}{\mathcal{I}_{K}^{\mathrm{cm}}(\alpha)},
\label{eq:QCRI}
\end{equation}where $\mathcal{O}$ is defined in Eq.~\eqref{eq:observable_def}, and $\braket{\mathcal{O}}_{\alpha}\equiv\braket{\Psi_{K}(\alpha)|\mathcal{O}|\Psi_{K}(\alpha)}$ and $\sqrbr{\mathcal{O}}_{\alpha}\equiv\sqrt{\braket{\mathcal{O}^{2}}_{\alpha}-\braket{\mathcal{O}}_{\alpha}^{2}}$. When we only consider the first passage time observable $\mathcal{O}_f$, the left hand side of Eq.~\eqref{eq:QCRI} becomes 
\begin{equation}
\frac{\sqrbr{\mathcal{O}_{f}}_{\alpha}^{2}}{\left(\partial_{\alpha}\braket{\mathcal{O}_{f}}_{\alpha}\right)^{2}}=\frac{\frac{1}{(1+\alpha)^{2}}\sqrbr{\mathcal{O}_{f}}^{2}}{\left(\partial_{\alpha}\frac{1}{1+\alpha}\braket{\mathcal{O}_{f}}\right)^{2}}=(1+\alpha)^{2}\frac{\sqrbr{\mathcal{O}_{f}}^{2}}{\braket{\mathcal{O}_{f}}^{2}}.
\label{eq:CRI_lhs_scaling}
\end{equation}Using Eqs.~\eqref{eq:IC_le_IQ} and \eqref{eq:CRI_lhs_scaling}, we obtain Eq.~\eqref{eq:alpha_main_result}.

\end{document}